\documentclass[angeod,manuscript]{copernicus}  

\nolinenumbers
\usepackage{amsmath}
\usepackage{color}
\usepackage{babel}
\usepackage[T1]{fontenc}

\usepackage{natbib}

\begin{document}

\title{Seeds for collisionless reconnection
}

\author[1,3]{Rudolf A. Treumann}
\author[2]{Wolfgang Baumjohann}
\affil[1]{International Space Science Institute, Bern, Switzerland}
\affil[2]{Space Research Institute, Austrian Academy of Sciences, Graz, Austria}
\affil[3]{Geophysics Department, Ludwig-Maximilians-University Munich, Germany\\

\emph{Correspondence to}: Wolfgang.Baumjohann@oeaw.ac.at
}

\runningtitle{Reconnection theory}

\runningauthor{R. A. Treumann \& W. Baumjohann}

\received{ }
\pubdiscuss{ } 
\revised{ }
\accepted{ }
\published{ }


\firstpage{1}

\maketitle

\begin{abstract}
A stationary, two-dimensional, and non-driven antiparallel magnetic field configuration is considered consisting of a central current film, flanked by two non-magnetic domains of electron inertial scale, resembling the Josephson model.    Landau-Ginzburg conditions apply at the boundaries to the oppositely directed external magnetic fields, whose sources are located at infinity. On the microscopic level sufficiently large magnetic islands and X points form from exchange of small numbers of elementary magnetic fluxes without reference to any  plasma instabilities which on these scales are purely electrostatic. These may serve as seeds to ignite large-scale collisionless reconnection.  

\end{abstract}
\section{Introduction}
Magnetic reconnection is believed to be a fundamental process of energy conversion in both collisional (resistive) and collisionless (dilute high-temperature) plasmas. Collisional reconnection is a slow general diffusion process \citep[for transport theory cf., e.g.,][for transport theory in plasma]{huang1987,krall1973} in which the plasma diffuses across a given locally anti-parallel magnetic field configuration under partial pressure differences, and mixes, while the friction converts some amount of magnetic energy into heat. Its physics is very well known since  \citet{einstein1905}. Looked at from the magnetic point of view the magnetic field diffuses, transports the frozen particle component and converts a fraction of  magnetic into kinetic energy \citep{zweibel2009,yamada2010,baumjohann1996}. Collisionless reconnection, on the other hand, is not as well understood. It is usually considered  from the magnetic point of view   as spontaneous merging of anti-parallel magnetic fields (field lines) in the reconnection X point, which meet and annihilate thereby releasing the stored magnetic energy. Its resulting side effects \citep[cf., e.g.,][for a more recent review]{treumann2015} have been extensively studied by simulations. Still, the complex physics is incomplete leaving work for several more decades and generations of researchers.

\subsection{Current sheets and flux elements}
There are two subtle points about reconnection which should be clarified. First, reconnection speaks about annihilation of field lines \citep[cf., e.g.,][for an early review]{vasyliunas1975}. However, magnetic field lines are no physical objects. They are the non-material lines of force of the magnetic field $\vec{B}$  indicating its direction,  just providing a visualisation of the magnetic field structure. Themselves they lack any volume and proper dynamics. What takes part in the physics of reconnection are magnetic flux tubes of certain volume, flux content, stored magnetic energy and defined as the surface integrals of the magnetic field over the surface $\vec{F}$:
\begin{equation}
\Phi=\int \vec{B}\cdot d\vec{F}
\end{equation}
Reconnection is about the change of magnetic flux per time unit $\dot\Phi (t)$ when fluxes of opposing signs come into contact. This is given by the well-known line-integral of the electric-induction field around the reconnecting magnetic flux-tubes
\begin{equation}
\dot\Phi(t)=-\oint \vec{E}\cdot d\vec{s}
\end{equation}
What is exchanged and modified is the magnetic flux or, from the electric point of view, the finite electric potential around the contour enclosing the site of the two reconnecting flux tubes containing fluxes of opposite signs. This contour is difficult to define, and the annihilation of fluxes thus poses a problem.

From a microscopic viewpoint, we know that the flux is quantised with flux quantum $\Phi_0=\pi\hbar/2e\approx 10^{-15}$ Vs, a natural constant. Hence magnetic fluxes can be annihilated only in entire numbers of $\Phi_0$. At high temperatures the exact number of annihilated flux elements is not important. What is important is the average number $\langle N\rangle_\phi$ which takes part in the merging and annihilation. There always remains a number 
\begin{equation}
N_\phi=\frac{\Phi}{\Phi_0}-\langle N\rangle_\phi
\end{equation}
 of unreconnected fluxes. Since the lowest Landau level is $\frac{1}{2}\hbar\omega_0$, a flux element corresponds to the elementary flux tube radius $r_0=\sqrt{\Phi_0/\pi B}$.  From this a flux tube (or field line) of a $B= 1$ nT field has radius $r_{ce}\approx 10^3 r_0$ and contains $\langle N\rangle_\phi=\Phi/\Phi_0\approx 10^6$ flux quanta. The above equation then says that in the reconnection process the number of elementary flux quanta exchanged (annihilated, their magnetic energy content converted into other forms of energy) per unit time amounts to
\begin{equation}
\dot{\langle N\rangle}_\phi = -\Phi_0^{-1}\oint \vec{E}\cdot d\vec{s}\approx 2\pi r_{ce} \Phi_0^{-1}E_\phi =\Phi_0^{-1}\Delta U_B
\end{equation}
The right-hand side is an estimate for the reconnection potential $\Delta U_B$. If it can be measured, it gives the reconnected flux numbers. Thus the question arises whether or not we can say anything on the number of such flux quanta in reconnection. 

\begin{figure}[t!]
\centerline{\includegraphics[width=0.5\textwidth,clip=]{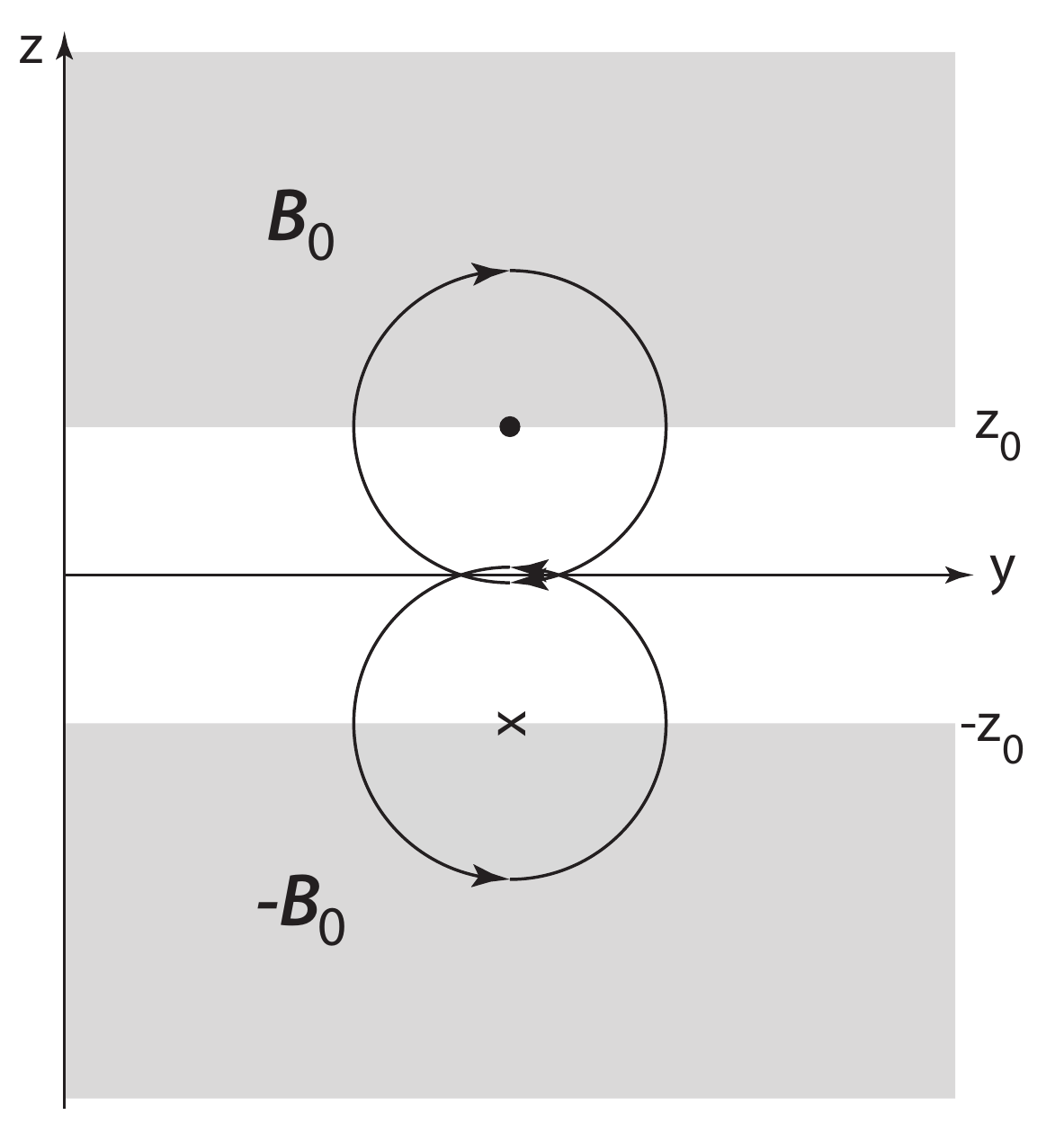}}
\caption{Schematic of generation of a current film in $y$ by the higher energy gyrating electrons of sufficiently large gyroradii around the last field line in the external mirror symmetric magnetic field $\pm\vec{B}_0$ caused by close overlap of cyclotron orbits. The currents in $-z_0<z<z_0$ due to electrons of smaller and larger gyroradii are comparably weak. The overlap is strongest near $z=0$ causing a current film.} \label{fig-elcurr}
\end{figure}

In the following we develop a primitive model in which we will be capable of illuminating the dynamics of the magnetic flux in stationary reconnection. 

Under stationary reconnection we understand that two regions of antiparallel fields come into contact without any dynamic inflow from the two sides. This is the state usually known as a current sheet in plasma and the investigation of its equilibrium respectively its stability. Stationary states of current sheets have been considered   in MHD \citep[cf., e.g.,][for general MHD reconnection in current sheets]{syrovatskii1971,biskamp2000} and  in kinetic theory \citep{harris1962} under the assumption that the entire field configuration is due to a broad current sheet. 

Our model is a modification of such models by dropping the MHD fluid assumption and including the electron inertial scale. In such an approach the  assumption is that the internal region of size $|z_0|\lesssim r_{ce}$ is essentially field free because the electrons are non-magnetic here and thus do not transport any magnetic field. Magnetic fields in the collisionless case can penetrate into this region only up to the electron skin depth $\lambda_e=c/\omega_e$, with $\omega_e=e\sqrt{N/\epsilon_0m_e}$ the electron plasma frequency based on the mean plasma number density $N$. This leads to a substantially different physics providing some insight into the reconnection process. We do not consider the case of driven reconnection with forced field and plasma inflow from both sides into the contact region.

The second problem is that reconnection must start in some way. Usually, in simulations, it is ignited by imposing a seed X point and observing the evolution of reconnection. A number of instabilities have been called for like the tearing mode which should start from thermal fluctuations in a magnetised plasma. These do not apply here in the non-magnetic central part. In the following we propose how a small scale X point forms when a small excess or lack of magnetic flux exists somewhere in the current layer.  This X point respectively the related plasmoid may then serve as the initial ignition of reconnection. Since very large numbers of fluxes are involved in reconnection, it is quite clear that the uncertainty of their number easily leaves space for fluctuations in the number of flux elements such that $N_\phi$ excess flux elements will become active in the way shown below. 

\section{The model}
We consider a classical modified Josephson junction \citep{josephson1962,josephson1964,josephson1974} model of the reconnection layer. It consists of a very narrow current layer or film of thickness $d$ in the centre whose current is not responsible for the large-scale antiparallel field configuration. This differs from common settings of reconnection where it is assumed that the antiparallel magnetic fields $\pm B_{0x}$ are produced by such a current layer. The assumption here is that the primary sources of these fields are currents flowing far away in space at $z\to\pm\infty$, and their nature is of no interest here nor in the following. These antiparallel fields are separated by a no-external field region of  thickness $\Delta z=\pm 2z_0$. The case of a vacuum is of no interest in reconnection as the antiparallel fields in vacuum have arbitrary possibility to rearrange. We therefore consider the opposite option in which the no-field region is an ideal conductor. It prevents the external field from penetrating it deeper than one skin depth $\lambda_e$. 

A no-field region arises because the external magnetic field is screened, the case realised for an ideal conductor. It does not contain an initial magnetic field nor magnetic flux except for the field which enters over the electron skin depth $\lambda_e=c/\omega_e$. Since electrons have gyro radii according to their energies, this region contains a weak distributed current caused by the antiparallel gyrations in the opposite external fields which we neglect here as it is of no interest for our purposes. However, this current maximises in the centre of the no-field layer which corresponds to a current film as shown in Figure \ref{fig-elcurr}.

Let this current suddenly, at some location $z=0, x=x_0$, strengthen or weaken from some statistical thermal fluctuation over some finite short lengths along $y$ and $x$. This weakening or strengthening can be understood as the sudden appearance of a number $N_\phi$ of flux elements $\Phi_0$ in the origin \citep{pearl1966,clem1974}, which is the modification introduced on the Josephson junction in order to adopt it to the current film and ultimately also reconnection.  Then the question arises what their effect would be on the overall field configuration. 

\subsection{Flux elements and current vortex}
We are dealing with a classical physics problem. However, reconnection implies the exchange and annihilation of a substantial number of flux elements. This will never be complete, causing some flux elements $N_\phi$ locally to be either in excess or lacking. This positive or negative accumulation of $N_\phi$ flux elements in the origin is a microscopic quantum effect which cannot be neglected a priori because it will generate a macroscopic effect in the magnetic field which reconnection will amplify. We shall show that the presence of even such a small number $N_\phi$ of flux elements will, classically, produce seeds for reconnection which may become of the vertical size $z_0$ of the no-field region. 

A local accumulation of $N_\phi$ excess flux elements implies the presence of a related sheet current $\vec{j}$ which forms a two-dimensional vortex closing in itself. The magnetic field of this sheet current is described by its vector potential $\vec{A}$. The ideally conducting region between the external fields and the current screens its induced field with screening length $\lambda$. In general $\lambda\neq\lambda_e$ may differ from the electron inertial length, depending on the number of electrons taking part in the screening. Accounting for this screening effect by maintenance of the number of flux elements can be done by reference to the current in Ginzburg-Landau theory \citep{landau1941,ginzburg1950,ginzburg1955}, when in our classical theory neglecting the appearing quantum-current terms while keeping the gauge contribution of the flux elements. This corresponds to the London-Josephson approximation. We are then left with the current density
\begin{equation}
\vec{j}=-\frac{1}{\mu_0\lambda^2}\Big(\vec{A}-\frac{\Phi_0}{2\pi}\nabla S\Big)
\end{equation}
where $S=N_\phi\theta$ is the dimensionless phase of the wave function $\psi$ of the participating electrons with density $|\psi|^2/N=\alpha$, the order parameter in Landau-Ginzburg and Josephson theory, which are involved. The gauge function in the last term counts the number of magnetic flux elements as they are the ingredient that contributes to the current vortex. 

Since the $N_\phi$ flux elements are  localised in $x=0$, they all together represent a magnetic island in the centre of the sheet and, in addition to the tangential magnetic field components, also produce a vertical magnetic component $B_z$ that extends over some distance into the no-field region and forms a small plasmoid. In our classical case which maintains these flux elements, because for any realistic magnetic field in space they come in susceptible numbers, the gauge phase itself appears only as an intermediate step and drops out later when performing the gradient. Physically it is the gauge of the vector potential which is determined only up to the gradient of an arbitrary scalar while microscopically connecting to the number of involved flux elements. The length $\lambda$ in this interpretation is the London penetration length scale
\begin{equation}
\lambda=c/\sqrt{\alpha}\:\omega_e
\end{equation}
Note again that we are  dealing with the classical case, having dropped all quantum terms except the phase in the gauge in order to keep the connection to the number of un-annihilated magnetic fluxes involved. They and the field are the physical quantities which participate, not any field lines which lack any physical substance.

The current $\vec{j}$ is restricted to the narrow film in the centre, and we need to consider only the height integrated current density integrated in $z$ over the  thickness $d$ of the film related to the flux elements in the  current sheet
\begin{equation}\label{eq-j}
\vec{\mathcal{J}}=-\frac{1}{\mu_0\Lambda}\Big(\vec{A}-\frac{\Phi_0}{2\pi}\nabla S\Big), \qquad \Lambda=\frac{\lambda^2}{d}
\end{equation}
From now on we deal with the region $z\neq0$ only. Outside the current $\vec{\mathcal{J}}$ there is no current flow, and thus the vector potential obeys the Laplace equation
\begin{equation}
\nabla^2\vec{A}=0, \qquad z\neq 0
\end{equation}
The total vector potential $\vec{A}$ in the region $0<|z|<|z_0|$ is the sum of the two vector potentials
\begin{equation}
\vec{A}=\vec{A}_0+\vec{A}_\theta, \quad 0<|z|<|z_0|
\end{equation}
with $\vec{A}_0=A_{0y}$ the external field potential given by
\begin{equation}
A_{0y}(z)=\pm \lambda_eB_0\,e^{-|z_0-z|/\lambda_e} +\nabla_y a(x,y),\quad |z|\leq|z_0|
\end{equation}
with arbitrary potential field $a(x,y)$. 

\subsection{Vector potential}
Since the flux elements are localised at the origin, it is  convenient to assume cylindrical coordinates $x=\rho\cos\theta,y=\rho\sin\theta,z$ and $S=N_\phi\theta$. In principle, we may later specify to a distribution of flux elements over the current sheet. It is however more important to investigate what the effect of a single accumulation is. For a distribution the solution obtained could serve as a Greens function.

With this specification the problem for $\vec{A}_\theta=A_\theta(\rho,z)\vec{\theta}$ in the current free region and mirror symmetry in $z$ to both sides of the current sheet is cylindrically symmetric in the coordinates $\rho,\theta,z$, obeying the Laplace equation
\begin{equation}
\Big(\frac{\partial^2}{\partial\rho^2}+\frac{1}{\rho}\frac{\partial}{\partial\rho}-\frac{l^2}{\rho^2}+\frac{\partial^2}{\partial z^2}\Big)A_\theta =0, \qquad z\neq 0
\end{equation}
Since the ideal conductor only permits the field to penetrate over the integrated London skin depth $\Lambda$, the $z$-dependence produces an exponential factor $Z(z)\sim\exp(-k|z|)$. Separating the above equation according to $A_\theta=R(\rho)\Theta(\theta)Z(z)$ with $l=1$ for the assumed circular symmetry, we  arrive at Bessel's differential equation 
\begin{equation}
\rho^2 R''(\rho)+\rho R'(\rho)-\Big(1-k^2\rho^2\Big)R(\rho)=0, \qquad z\neq0
\end{equation}
for $R(\rho)$ with  solution $R(\rho)=J_1(k\rho)$. [If the circular symmetry would be violated, we had to retain the sum over all $l$ and include the trigonometric functions $e^{\pm il\theta}$.] This allows to write the general solution for the external vector potential of the current sheet in the no-field region as
\begin{equation}\label{eq-atheta}
A_{\theta}(\rho,z)=\int_0^\infty dk\:J_1(k\rho)\Big[b_k\, e^{-k|z|}+c_k\, e^{k|z|}\Big], \quad z\neq 0
\end{equation}
where $b_k, c_k$ are the Bessel expansion coefficients as functions of the Bessel wavenumber $k$. They account for the presence of  outgoing and incoming potentials to the extent that the screening skin effect is warranted. 

\subsection{Josephson boundary conditions}
For $\vec{A}_\theta$ we have to satisfy boundary conditions at $z=0$ and $|z|=|z_0|$. The first boundary condition implies that the $\rho$ component of the magnetic field is discontinuous at $z=0$ yielding \begin{equation}\label{eq-brho}
{\mathcal{J}}_\theta(\rho,0)=\frac{1}{\mu_0}\big[B_\rho(\rho,0_+)-B_\rho(\rho,0_-)\big]
\end{equation}
where $B_\rho=-\partial_z A_\theta,\: B_\theta=0$. On the other hand we have from (\ref{eq-j}) at $z=0$ and with 
\begin{equation}
\nabla S=\frac{1}{\rho}\frac{\partial S}{\partial\theta}=\frac{N_\phi}{\rho}
\end{equation}
 for the same current
\begin{equation}\label{eq-jtheta}
{\mathcal{J}}_\theta(\rho,0)=-\frac{1}{\mu_0\Lambda}\Big[A_\theta(\rho,0)-\frac{N_\phi\Phi_0}{2\pi\rho}\Big]
\end{equation}
 

\begin{figure}[t!]
\centerline{\includegraphics[width=0.5\textwidth,clip=]{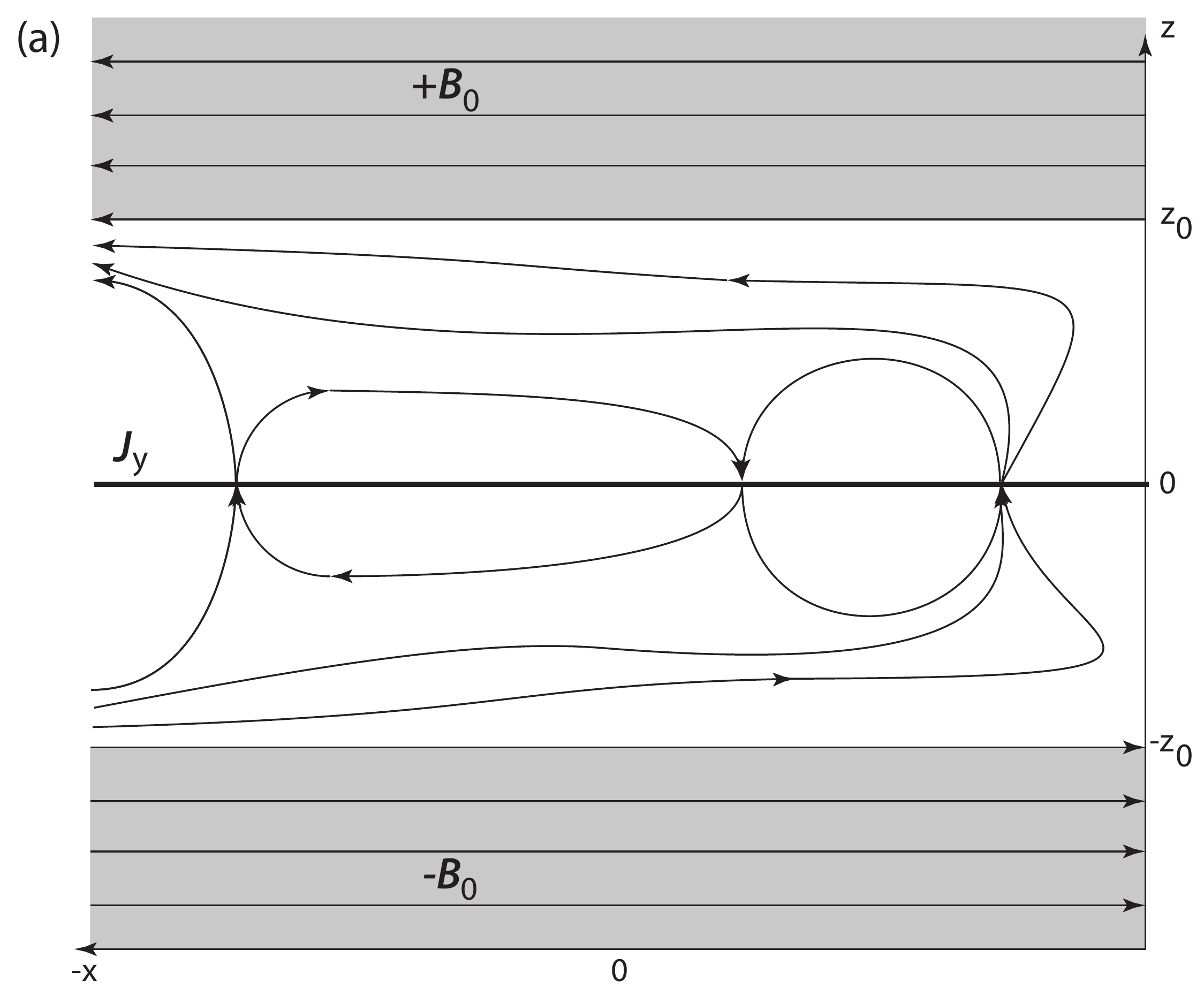}}
\centerline{\includegraphics[width=0.5\textwidth,clip=]{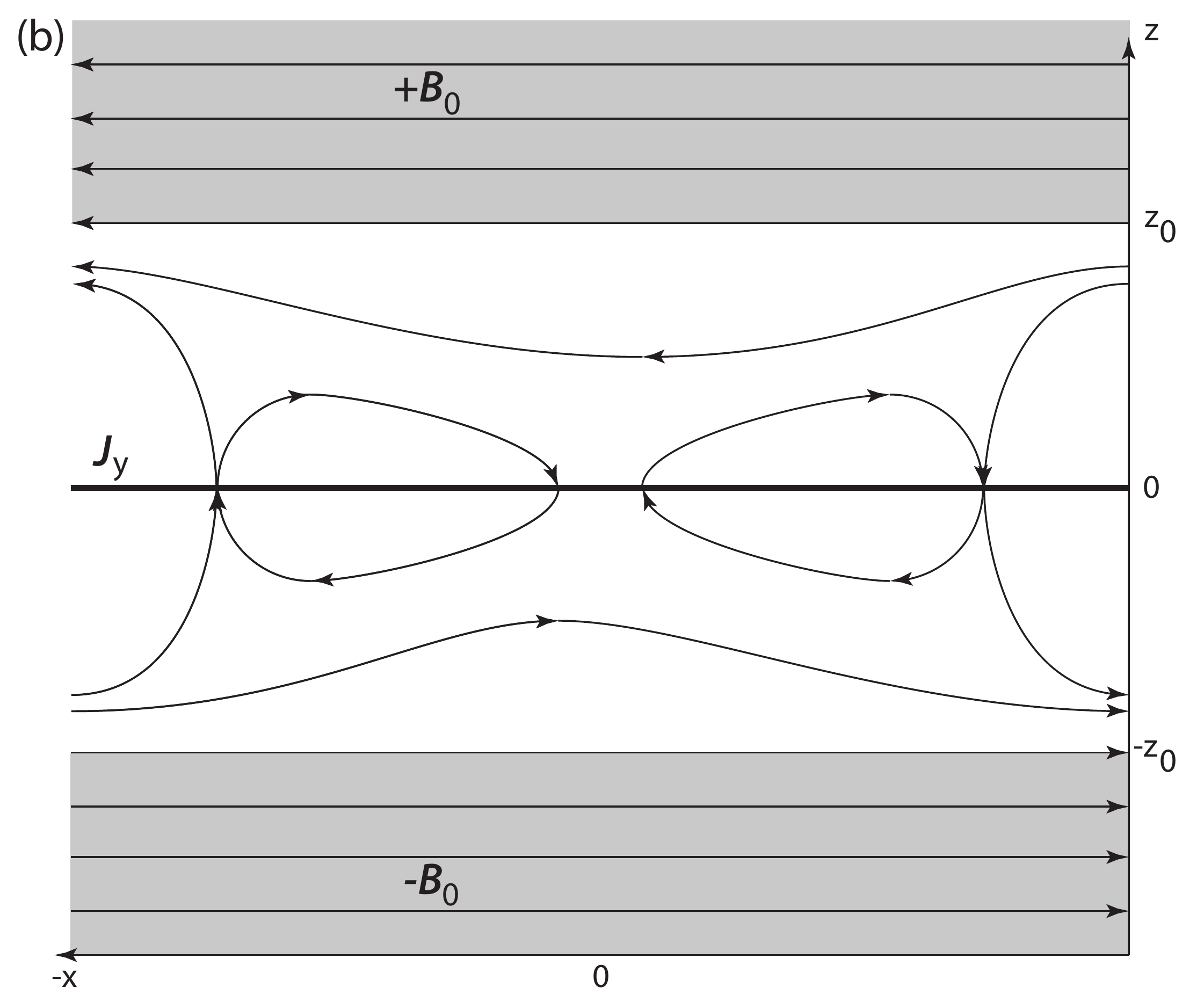}}
\caption{A sketch of the field configuration in the region $|z|<|z_0|$ with superimposed external field and the field resulting from a localised accumulation of elementary fluxes in the thin central current sheet. Electrons are non-magnetic here. The current is out of the plane. Hence, there is no dynamics, and the fields can freely rearrange. The field configuration produced by a single distortion of the central current sheet is antisymmetric and possesses a typical structure. At large distances along the current layer the external field dominates. In the near zone the distortion causes a more complicated field configuration. \emph{Upper part a}: Two parallel (northward) flux accumulations along the $x$ axis in the $(x,z)$ plane at $y=0$. This case is unrealistic because the parallel fluxes will readily attract and merge into one, as is seen from the configuration of their combined fields. The X point in the middle will become squashed. \emph{Lower part b}: Two antiparallel flux accumulations. In both cases several X points and magnetic islands arise.  } \label{fig-single}
\end{figure}


Expressing the field components in (\ref{eq-brho}) through the derivative of the vector potential (\ref{eq-atheta}) and combining  with (\ref{eq-jtheta}) for $\mathcal{J}_\theta(\rho,0)$, and (\ref{eq-j}), yields an equation
\begin{equation}
\int_0^\infty dk\:\Big[b_k(2k\Lambda+1)-c_k(2k\Lambda-1)-\frac{N_\phi\Phi_0}{2\pi}\Big]J_1(k\rho)=0
\end{equation}
where we used the identity $\int_0^\infty dxJ_1(x)=1$. This gives a relation between the Fourier coefficients
\begin{equation}
b_k(2k\Lambda+1)-c_k(2k\Lambda-1)=\frac{N_\phi\Phi_0}{2\pi}
\end{equation}
A second relation is obtained from the requirement of the vanishing of the tangential component $B_\rho(\pm z_0)=-\partial_z A_\theta(\rho,z)|_{\pm z_0}=0$ at the boundary $|z_0|$ of the no-field domain
\begin{equation}
b_ke^{\mp kz_0}-c_ke^{\pm kz_0}=0
\end{equation}
Solving for the Bessel coefficients in the upper domain $z_0>0$ yields
\begin{equation}
(b_k,c_k)=\frac{N_\phi\Phi_0}{4\pi}\frac{\exp(\pm kz_0)}{2k\Lambda\sinh(kz_0)+\cosh(kz_0)}
\end{equation}
The same expressions hold at the lower boundary with redefinition of the signs on $z_0$. 

\subsection{Field and current}

We introduce the dimensionless variables  $\kappa=2k\Lambda, \xi=\rho/2\Lambda, \zeta=z/2\Lambda$. Then we are in the position to write down the expressions for the vector potential, the two field components, and the sheet current strength in the region $\rho>0, |z|\leq|z_0|$:
\begin{eqnarray}
A_\theta(\rho,z)\!\!\!\!\!&=&\!\!\!\!\!\frac{N_\phi\Phi_0}{4\pi\Lambda}\int_0^\infty\frac{d\kappa\,J_1(\kappa\xi)\cosh \kappa |\zeta-\zeta_0|}{\kappa\sinh(\kappa|\zeta_0|)+\cosh(\kappa|\zeta_0|)}\\
B_\rho(\rho,z)\!\!\!\!\!&=\pm&\!\!\!\!\!\frac{N_\phi\Phi_0}{8\pi\Lambda^2}\int_0^\infty\frac{\kappa d\kappa\,J_1(\kappa\xi)\sinh \kappa|\zeta-\zeta_0|}{\kappa\sinh(\kappa|\zeta_0|)+\cosh(\kappa|\zeta_0|)}\\
B_z(\rho,z)\!\!\!\!\!&=&\!\!\!\!\!\frac{N_\phi\Phi_0}{8\pi\Lambda^2}\int_0^\infty\frac{\kappa d\kappa\,J'_1(\kappa\xi)\cosh \kappa|\zeta-\zeta_0|}{\kappa\sinh(\kappa|\zeta_0|)+\cosh(\kappa|\zeta_0|)}\\
\mathcal{J}_\theta(\rho,0)\!\!\!\!\!&=&\!\!\!\!\!\frac{N_\phi\Phi_0}{4\pi\mu_0\Lambda^2}\int_0^\infty\frac{\kappa d\kappa\,J_1(\kappa\xi)}{\kappa+\coth(\kappa|\zeta_0|)}
\end{eqnarray}
where $J'_1(x)\equiv dJ_1(x)/dx=J_0(x)-J_1(x)/x$. Unfortunately, none of these integrals can be given in closed form. Only in the case $|z_0|\to\infty$, which is of no interest here, can a closed form be obtained for the sheet current:
\begin{equation}
\mathcal{J}_\theta(\rho,0)=\frac{N_\phi\Phi_0}{4\pi\mu_0\Lambda^2}\int_0^\infty \:\frac{\kappa\,d\kappa J_1(\kappa\xi)}{\kappa +1}
\end{equation}
An equation like this has a well known solution in high temperature multi-layer superconductivity \citep{pearl1966,clem1974,clem1991} which can be constructed with the help of a Hankel transformation  \citep[see][pp. 685 \& 983]{gradshteyn1965} yielding after some manipulations for $|z_0|\to\infty$
\begin{equation}
\mathcal{J}_\theta(\rho, 0)= \frac{N_\phi\Phi_0}{2\mu_0\Lambda^2}\Big[\mathbf{H}_1(\rho/2\Lambda)-N_1(\rho/2\Lambda)-\frac{2}{\pi}\Big]
\end{equation}
Here $\mathbf{H}_1(x), N_1(x)$ are the Struve and Neumann (second kind Bessel) functions whose small and asymptotic argument expansions are well known but do not play any role in our strongly modified case.

Before discussing the properties of the solution in our case we observe that the magnetic field possesses a non-vanishing $z$ component which even remains finite at the transition from the no-field domain into the external field, while being continuous across the boundary $|z_0|$. This component superimposes onto the external field and causes a magnetic deformation. Though the structure of the field is not obvious we observe that $B_z(\zeta)$ has same sign above and below the current sheet, but varies with $\rho$. The localised compound of flux elements thus has  properties similar to a localised dipole though the field is more complicated not having the simple dipolar geometry. 

\subsection{Properties of field}
The value of $B_z(\rho,z)$ at $z=|z_0|$ is proportional to the number of flux elements. At $\rho=0$ above the origin we have $\lim_{x\to0}J'_1(x)/x=\frac{1}{2}$, for instance. Then one has
\begin{eqnarray}
B_z(0,\zeta_0)&=&\frac{N_\phi\Phi_0}{16\pi\Lambda^2|\zeta_0|}\int_0^\infty\frac{\kappa' d\kappa'}{\kappa'\sinh(\kappa')+|\zeta_0|\cosh(\kappa')}\nonumber\\
& =&\frac{N_\phi\Phi_0}{16\pi\Lambda^2|\zeta_0|}R(z_0/2\Lambda)
\end{eqnarray}
The integral is just a number $R(|z_0|)$. Hence the $z$ component of the magnetic field is a measure of the number $N_\phi$ of flux elements involved in the distortion of the current sheet.

The magnetic flux is twofold. There is a circular flux $\Phi_\theta$ which closes in itself. In addition there is the flux in $z\ (\zeta)$ direction. This has a typical dependence on $\rho$ respectively $\xi$. It is obtained integrating $B_z$ over the circular cross section $2\pi\xi d\xi$ yielding
\begin{eqnarray}
\frac{\Phi_z(\xi)}{Q(N_\phi)}\!\!\!\!\!&=&\!\!\!\!\!\int_0^\infty\frac{d\kappa\,[\kappa\xi J_1(\kappa\xi)+J_0(\kappa\xi)-1]\cosh \kappa(\zeta-\zeta_0)}{\kappa[\kappa\sinh(\kappa|\zeta_0|)+\cosh(\kappa|\zeta_0|)]}\nonumber\\[-1.5ex]
&&\\[-1.5ex]
Q(N_\phi)\!\!\!\!\!&=&\frac{N_\phi\Phi_0}{4\Lambda^2}\nonumber
\end{eqnarray}
The apparent singularity at $\kappa=0$ is compensated by the behaviour of the nominator. This can be shown applying l'Hospital's rule.   

Both field components decay with increasing radius $\rho$ respectively $\xi$ from the origin approximately $\sim \xi^{-1/2}$ causing the field to become weak at large distance from the origin. At those distances the external field dominates with a small contribution of the residual induced sheet current field which causes a weak amplification of the external field close to the sheet. 

The total field in $|z_0-z|$ is the sum of the above field components and the external field 
\begin{equation}
B_{0x}\vec{x}=\pm B_0\vec{x}\exp\Big(-\frac{\Lambda}{\lambda_e}|\zeta_0-\zeta|\Big)
\end{equation}
where $\vec{x}=\vec{\rho}\cos\theta-\vec{\theta}\sin\theta, \ \vec{y}=\vec{\rho}\sin\theta+\vec{\theta}\cos\theta$. Depending on the ratio $\Lambda/\lambda_e$ the external field is more or less damped in $\zeta$. The geometrical superposition of both fields determines the field structure in the intermediate region outside $z=0$. In the $(x,z)$ plane one has $\theta=0, \xi=x/2\Lambda$, and $B_\rho=B_x$.

\subsection{Field line geometry}
Though the above integrals for the field components cannot be solved analytically and one has to turn to numerical calculations, it is possible to obtain an impression on the structure of the contributions to the total magnetic field above and below the current sheet. The equation of the lines of force of the elementary contributions is, in the azimuthally symmetrical case under consideration, obtained from 
\begin{equation}
\frac{d\zeta}{d\xi} = \frac{B_z}{B_\rho}
\end{equation}
Integration with respect to $\xi$ can be performed in the limits of small $\rho$ near the origin, and at large radial distance as $\zeta(\xi)$ when using the appropriate expansions of $J_1(x), J'_1(x)$. Near the origin $\zeta=0,\xi\ll 1$, and with $f(\zeta_0)$ the ratio of the remaining $\kappa$ integrals at $\zeta=0$, this gives
\begin{equation}
\zeta(\xi)\sim f(\zeta_0)\,\xi^2, \qquad \xi\ll1
\end{equation}
a quadratic dependence on $\xi$ of the magnetic force lines in the plane $(\zeta,\xi)$. Vice versa, at large radii $\xi\gg 1$ and $\zeta\sim\zeta_0$, using the asymptotic expansions of the Bessel function and its derivative yields 
\begin{equation}
\zeta(\xi)\sim g(\zeta_0)\sqrt{\xi},\qquad \xi\gg1, \ |\zeta_0-\zeta|\ll1
\end{equation}
where use has been made of the behaviour of the hyperbolic functions at small argument for $|\zeta-\zeta_0|\ll 1$, and $g(\zeta_0)$ is the ratio of the remaining $\kappa$ integrals. This field is to be overlaid on the external field. Because all particles are nonmagnetic, there is no dynamics in the region $|z_0|>|z|$. The fields, being subject to their stresses, can freely rearrange (except for the reduced light speed in the dielectric medium). The superposition generates a typical structure for this kind of current layers given under symmetric conditions in Figure \ref{fig-single} for the two cases of two parallel (northward) and two antiparallel flux accumulations. In both cases the magnetic field assumes a fairly complicated structure. In the transition region between the two external fields magnetic islands arise as well as extended field-free regions or magnetic holes. These may be understood as either holes or X points though there is no dynamics of the contained electrons involved.

\section{Discussion}
The philosophy in this note was that in reconnection on the elementary level one does not deal with magnetic field lines but flux tubes which have finite size and appear in quanta. On the large scale this is not felt, but in reconnection never all flux elements are exchanged due to thermal fluctuations and uncertainty on the microscopic level. Since in the annihilation of fairly strong magnetic fields very many elementary flux quanta are involved, the fluctuating part plays an essential role. We have shown that in a setting where no initial current layer exists, the case which becomes real when two oppositely directed magnetic fields approach each other, as is the case for instance in magnetic fields of different polarity in the solar wind whose sources are located far away in the solar corona, then there is no internal current layer present as long as the distance between the oppositely directed fields exceeds the electron gyroradii. Over this distance the fields can penetrate only for an electron skin depth $\lambda_e$ being exponentially squeezed. Any small thermal fluctuation in the centre then causes an excess or lack of magnetic flux elements. These become sources of small-scale magnetic islands and X points which subsequently can serve as seeds for igniting reconnection. Both cases, excess or lack of flux elements produce such magnetic distortions in the separating domain. Since they superimpose on the external field, which penetrates over the skin depth, they provide the connection between the interior of the separating layer and the external fields, a combination which may subsequently serve as the necessary seeds to start reconnection. 

In this way, reconnection depends on the internal existence of flux elements and becomes the macroscopic effect of a well known quantum process. For a more complete theory the distributed weak electron current in $|z|<|z_0|$ should be taken into account. Moreover, the structure of the field suggests that for a substantially numerous accumulation of flux elements electron-inertial size magnetic holes could evolve which would be completely inactive as seen from the plasma physics view point. Though in pressure balance, there would not be any activity in their interior.

\begin{acknowledgement}
This work was part of a brief Visiting Scientist Programme at the International Space Science Institute Bern. We acknowledge the interest of the ISSI directorate as well as the generous hospitality of the ISSI staff, in particular the assistance of the librarians Andrea Fischer and Irmela Schweitzer, and the Systems Administrator Saliba F. Saliba. 
\end{acknowledgement}

\noindent\emph{Author contribution.} All authors contributed equally to this paper. 

\noindent\emph{Data availability.} No data sets were used in this article. 

\noindent\emph{Competing interests.} The authors declare that they have no conflict of interest.


\end{document}